\documentclass[runningheads]{llncs}
\usepackage{graphicx}
\usepackage{subfigure}
\usepackage{amsfonts}
\usepackage[colorlinks]{hyperref}
\usepackage{marvosym}
\usepackage{floatrow}
\newfloatcommand{capbtabbox}{table}[][0.48\textwidth]
\begin{document}
\title{PHTrans: Parallelly Aggregating Global and Local Representations for Medical Image Segmentation}
\titlerunning{PHTrans: Parallelly Aggregating Global and Local Representations}

\author{Wentao Liu\inst{1}\and Tong Tian\inst{2}\and Weijin Xu\inst{1} \and Huihua Yang \inst{1,3(\textrm{\Letter})}  \and Xipeng Pan\inst{3,4} \and Songlin Yan\inst{1} \and Lemeng Wang\inst{1}}

\authorrunning{W. Liu et al.}

\institute{School of Artificial Intelligence, Beijing University of Posts and Telecommunications, Beijing 100876, China\\
\email{yhh@bupt.edu.cn}\\
\and State Key Laboratory of Structural Analysis for Industrial Equipment, School of Aeronautics and Astronautics, Dalian University of Technology, Dalian 116024, China\\
\and School of Computer Science and Information Security, Guilin University of Electronic Technology, Guilin 541004, China
\and Department of Radiology, Guangdong Provincial People’s Hospital, Guangdong Academy of Medical Sciences, Guangzhou 510080, China
}


%
\maketitle           
\begin{abstract}

The success of Transformer in computer vision has attracted increasing attention in the medical imaging community. Especially for medical image segmentation, many excellent hybrid architectures based on convolutional neural networks (CNNs) and Transformer have been presented and achieve impressive performance. However, most of these methods, which embed modular Transformer into CNNs, struggle to reach their full potential. In this paper, we propose a novel hybrid architecture for medical image segmentation called PHTrans, which parallelly hybridizes Transformer and CNN in main building blocks to produce hierarchical representations from global and local features and adaptively aggregate them, aiming to fully exploit their strengths to obtain better segmentation performance. Specifically, PHTrans follows the U-shaped encoder-decoder design and introduces the parallel hybird module in deep stages, where convolution blocks and the modified 3D Swin Transformer learn local features and global dependencies separately, then a sequence-to-volume operation unifies the dimensions of the outputs to achieve feature aggregation. Extensive experimental results on both Multi-Atlas Labeling Beyond the Cranial Vault and Automated Cardiac Diagnosis Challeng datasets corroborate its effectiveness, consistently outperforming state-of-the-art methods. 
The code is available at: \url{https://github.com/lseventeen/PHTrans}.

\end{abstract}
\section{Introduction}

Medical image segmentation aims to extract and quantify regions of interest in biological tissue/organ images, which are essential for disease diagnosis, preoperative planning, and intervention. Benefiting from the excellent representation learning ability of deep learning, convolutional neural networks (CNNs) have achieved tremendous success in medical image analysis. Many excellent network models (e.g., U-Net~\cite{unet}, 3D U-Net~\cite{3dunet} and Attention U-Net~\cite{ATT-UNet}) have emerged, constantly refreshing the upper limit of performance for various segmentation tasks. In spite of achieving extremely competitive results, CNN-based methods lack the ability to model long-range dependencies due to inherent inductive biases such as locality and translational equivariance. Several researchers have alleviated this problem by increasing the size of the convolution kernel~\cite{largekernel}, using atrous convolution~\cite{dilated}, and embedding self-attention mechanisms~\cite{non-local}. However, it cannot be fundamentally solved provided that the convolution operation remains at the heart of the network architecture.

\begin{figure}[tbp]
\centering
 \includegraphics[scale=0.55]{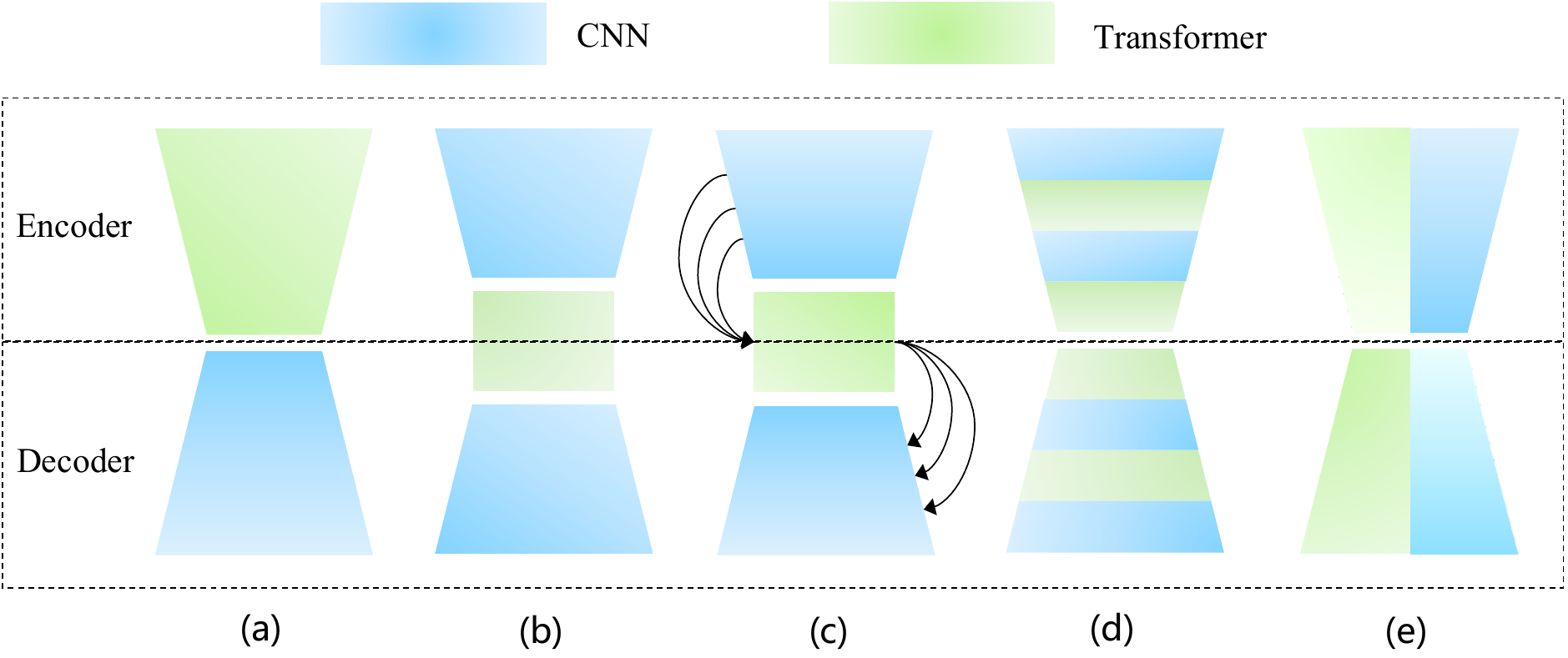}
\caption{Comparison of several hybrid architectures of Transformer and CNN. 
(Color figure online)} \label{fig1}
\end{figure}

Transformer~\cite{transformer}, relying purely on attention mechanisms to model global dependencies without any convolution operations, have emerged as an alternative architecture which has delivered better performance than CNNs in computer vision (CV) on the condition of being pre-trained on large-scale datasets. Among them, Vision Transformer (ViT)~\cite{ViT} splits images into a sequence of tokens and models their global relations with stacked Transformer blocks, which have revolutionized the CV field. Swin Transformer~\cite{Swin} can produce hierarchical feature representations with lower computational complexity in shiftable windows, achieving state-of-the-art performance in various CV tasks. However, medical image datasets are much smaller in magnitude than the datasets used by the pre-trained in the mentioned works (e.g., ImageNet-21k and JFT-300M) because medical images are not always available and require professional annotation. As a result, Transformer produces unsatisfactory performance in medical image segmentation. Meanwhile, many hybrid structures derived from the combination of CNN and Transformer have emerged, which offer the advantages of both and have gradually become a compromise solution for medical image segmentation without being pre-trained on large datasets.

We summarize several popular hybrid architectures based on Transformer and CNN in medical image segmentation. These hybrid architectures add Transformer into models with CNN as the backbone, or replace parts of the architecture's components. For example, UNETR~\cite{unetr} and Swin UNETR~\cite{swin-unet} used the encoder-decoder structure in which the encoder is composed of a cascade of blocks built with self-attention and multilayer perceptron, i.e., Transformer, while the decoder is stacked convolutional layers, see Fig.~\ref{fig1}(a). TransBTS~\cite{Transbts} and TransUNet~\cite{transunet} introduced a Transformer between the encoder and decoder composed of CNN, see Fig.~\ref{fig1}(b). MISSFormer~\cite{missformer} and CoTr~\cite{cotr} bridged all stages from encoder to decoder by Transformer instead of only adjacent stages, which captures the multi-scale global dependency, see Fig.~\ref{fig1}(c). In addition, nnFormer~\cite{nnformer} interleaved Transformer and convolution blocks into a hybrid model, where convolution encodes precise spatial information and self-attention captures global context, see Fig.~\ref{fig1}(d). From Fig.~\ref{fig1}, it can be seen that these architectures implement a serial combination of Transformer and CNN from a macro perspective. Nevertheless, in the serial combination, convolution and self-attention cannot run through the entire network architecture, making it difficult to continuously model local and global representations, so it does not fully exploit their potential.

In this paper, we propose a parallel hybrid Transformer (PHTrans) for medical image segmentation where the main building blocks consist of CNN and Swin Transformer to simultaneously aggregate global and local representations, see Fig.~\ref{fig1}(e). In PHTrans, we extend the standard Swin Transformer to a 3D version by extracting 3D patches that partition a volume and constructing 3D self-attention mechanisms. Given the hierarchical property of the Swin Transformer can conveniently leverage advanced techniques for dense prediction such as U-Net~\cite{Swin}, we followed the successful U-shaped architecture design and introduced a transformation operation of sequence and volume to achieve the parallel combination of Swin Transformer and CNN in a block. In contrast to serial hybrid architecture, PHTrans can independently construct hierarchical local and global representations and fuse them in each stage, fully exploiting the potential of CNN and the Transformer. Extensive experiments demonstrate the superiority of our method against other competing methods on various medical image segmentation tasks.

\section{Method}
\subsection{Overall Architecture}
\begin{figure}[t]
\centering
\includegraphics[scale=0.53]{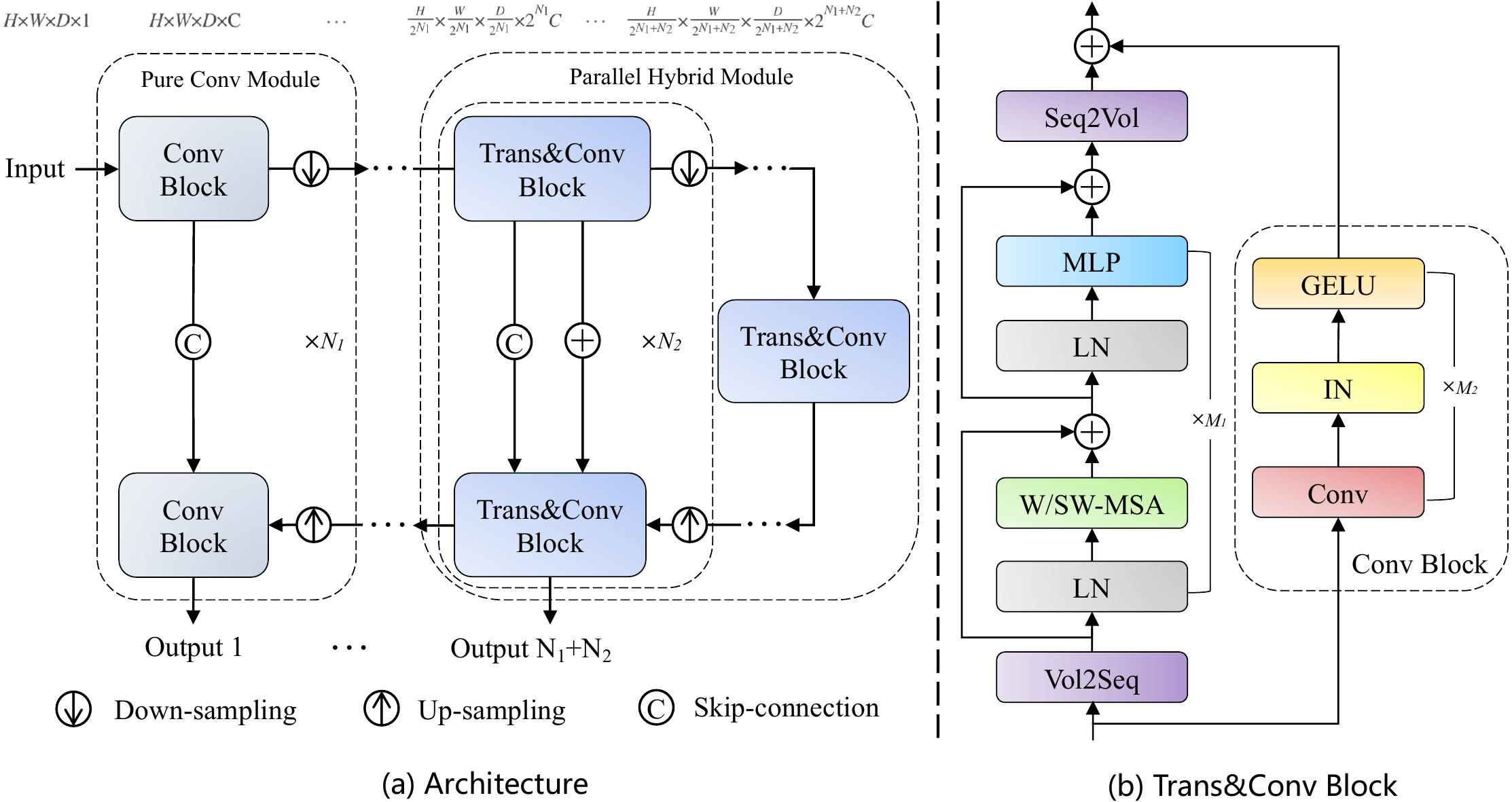}
\caption{(a) The architecture of PHTrans; (b) Parallel hybird block consisting of Transformer and convolution (Trans\&Conv block).  } \label{fig2}
\end{figure}

An overview of the PHTrans architecture is illustrated in Fig.~\ref{fig2}(a). PHTrans follows the U-shaped encoder and decoder design, which is mainly composed of pure convolution modules and parallel hybrid ones. Our original intention was to construct a completely hybrid architecture composed of Transformer and CNN, but due to the high computational complexity of the self-attention mechanism, Transformer cannot directly receive input with pixels serving as tokens. In our implementation, a cascade of convolution blocks and down-sampling operations are introduced to reduce the spatial size, which progressively extracts low-level features with high resolution to obtain fine spatial information. Similarly, these pure convolution modules are deployed in the decoder at the same stage to recover the original image dimension by up-sampling.

Given an input volume $x \in \mathbb{R}^{H\times W\times D}$, where $H$, $W$ and $D$ denote the height, width, and depth, respectively, we first utilize several pure convolution modules to obtain feature maps $f \in \mathbb{R}^{\frac{H}{2^{N_1}}\times \frac{W}{2^{N_1}}\times \frac{D}{2^{N_1}} \times 2^{N_1}C}$, where $N_1$ and $C$ denote the number of modules and base channels, respectively. Afterwards, parallel hybrid modules consisting of Transformer and CNN were applied to model the hierarchical representation from local and global feature. The procedure is repeated $N_2$ times with $\frac{H}{2^{N_1+N_2}}\times \frac{W}{2^{N_1+N_2}}\times \frac{D}{2^{N_1+N_2}}$ as the output resolutions and $2^{N_1+N_2}C$ as the channel number. Corresponding to the encoder, the symmetric decoder is similarly built based on pure convolution modules and parallel hybrid modules, and fuses semantic information from the encoder by skip-connection and addition operations. Furthermore, we use deep supervision at each stage of the decoder during the training, resulting in a total of $N_1+N_2$ outputs, where joint loss consisting of cross entropy and dice loss is applied. The architecture of PHTrans is straightforward and changeable, where the number of each module can be adjusted according to medical image segmentation tasks, i.e., $N_1$, $N_2$, $M_1$ and $M_2$. Among them, $M_1$ and $M_2$ are the numbers of Swin Transformer blocks and convolution blocks in the parallel hybrid module.

\subsection{Parallel Hybrid Module}
The parallel hybrid modules are deployed in the deep stages of PHTrans, where the Trans\&Conv block, as its heart, achieves hierarchical aggregation of local and global representations by CNN and Swin Transformer.

\subsubsection{Trans\&Conv block.}
The scale-reduced feature maps are fed into Swin Transformer (ST) blocks and convolution (Conv) blocks, respectively. We introduce Volume-to-Sequence (V2S) and Sequence-to-Volume (S2V) operations at the beginning and end of ST blocks, respectively, to implement the transform of volume and sequence, making it concordant with the dimensional space of the output that Conv blocks produce. Specifically, V2S is used to reshape the entire volume (3D image) into a sequence of 3D patches with a window size. S2V is the opposite operation. As shown in Fig.~\ref{fig2}(b), a ST block consists of a shifted window based multi-head self attention (MSA) module, followed by a 2-layer MLP with a GELU activation function in between. A LayerNorm (LN) layer is applied before each MSA module and each MLP, and a residual connection is applied after each module~\cite{Swin}. In $M_1$ successive ST blocks, the MSA with regular and shifted window configurations, i.e., W-MSA and SW-MSA, is alternately embedded into ST blocks to achieve cross-window connections while maintaining the efficient computation of non-overlapping windows.

For medical image segmentation, we modified the standard ST block into a 3D version, which computes self-attention within local 3D windows that are arranged to evenly partition the volume in a non-overlapping manner. Supposing $x \in \mathbb{R}^{H\times W\times S \times C}$ is the input of ST block, it would be first reshaped to $N\times L\times C$, where $N$ and $L = W_h\times W_w\times W_s$ denote the number and dimensionality of 3D windows, respectively. The self-attention in each head is calculated as:

\begin{equation}
Attention(Q,K,V) =SoftMax(\frac{QK^T}{\sqrt{d}}+B)V,
\end{equation}
where $Q,K,V \in \mathbb{R}^{L\times d}$ are the $query$, $key$ and $value$ matrices, $d$ is the $query/key$ dimension, and $B \in \mathbb{R}^{L\times L}$ is the relative position bias. We parameterize a smaller-sized bias matrix $\hat{B}\in\mathbb{R}^{(2W_h-1)\times(2W_w-1)\times(2W_s-1)}$, and values in $B$ are taken from $\hat{B}$.

The convolution blocks are repeated $M_2$ times with a $3\times 3\times 3$ convolutional layer, a GELU nonlinearity, and an instance normalization layer as a unit. The configuration of the convolution blocks is simple and flexible enough that any off-the-shelf convolutional network can be applied. Finally, we fuse the outputs of the ST blocks and Conv blocks by an addition operation. The computational procedure of the Trans\&Conv block in the encoder can be summarized as follows:
\begin{equation}
y_i =S2V(ST^{M_1}(V2S(x_{i-1})))+Conv^{M_2}(x_{i-1}),
\end{equation}
where $x_{i-1}$ is the down-sampling results of the encoder's ${i-1}^{th}$ stage. In the decoder, besides skip-connection, we supplement the context information from the encoder with an addition operation. Therefore, the Trans\&Conv block in the decoder can be formulated as:
\begin{equation}
z_i =S2V(ST^{M_1}(V2S(x_{i+1}+y_{i}))+Conv^{M_2}([x_{i+1},y_{i}]),
\end{equation}
where $x_{i+1}$ is the up-sampling results of the decoder's ${i+1}^{th}$ stage and $y_{i}$ is output of the encoder's ${i}^{th}$ stage.

\subsubsection{Down-sampling and Up-sampling.} 
The down-sampling contains a strided convolution operation and an instance normalization layer, where the channel number is halved and the spatial size is doubled. Similarly, the up-sampling is a strided deconvolution layer followed by an instance normalization layer, which doubles the number of feature map channels and halved the spatial size. The stride is generally set to 2 in all dimensions. However, when 3D medical images are anisotropic, the stride with respect to specific dimensions is set to 1. 

\section{Experiments}
\subsection{Dataset}

The Multi-Atlas Labeling Beyond the Cranial Vault (BCV)~\cite{BCV} includes 30 cases with 3779 axial abdominal clinical CT images. Similar to \cite{nnformer}, the dataset is split into 18 training samples and 12 testing samples. And the average Dice-Similarity coefficient (DSC) and average Hausdorff Distance (HD) are used as evaluation metrics to evaluate our method on 8 abdominal organs (aorta, gallbladder, spleen, left kidney, right kidney, liver, pancreas, spleen, stomach).

The Automated Cardiac Diagnosis Challenge (ACDC)~\cite{ACDC} dataset is collected from different patients using MRI scanners. For each patient's MR image, left ventricle (LV), right ventricle (RV), and myocardium (MYO) are labeled. Following \cite{nnformer}, 70 samples are divided into the training set, 10 samples are divided into validation set, and 20 samples are divided into test set. The average DSC is used to evaluate our method on this dataset.

\subsection{Implementation Details}

For a fair comparison, we employed the code framework of nnUNet\cite{nnunet} to evaluate the performance of PHTrans as the same as CoTr\cite{cotr} and nnFormer\cite{nnformer}. All experiments were performed under the default configuration of nnUNet. In PHTrans, we empirically set the hyper-parameters [$N_1$,$N_2$,$M_1$,$M_2$] to [2,4,2,2] and adopted the stride strategy of nnU-Net~\cite{nnunet} for down-sampling and up-sampling without elaborate design. Moreover, the base number of channels C is 24 and the numbers of heads of multi-head self-attention used in different encoder stages are [3,6,12,24]. For BCV and ACDC datasets, we set the size of 3D windows [$W_h$,$W_w$,$W_s$] to [3,6,6] and [2,8,7] in ST blocks, respectively. In the training stage, we randomly cropped sub-volumes of size 48×192×192 and 16×256×224 from BCV and ACDC datasets as the input, respectively. We implemented PHTrans under PyTorch 1.9 and conducted experiments with a single GeForce RTX 3090 GPU. 

\subsection{Results}
\begin{table}[tbp]
\centering
\caption{Results of segmentation on the BCV dataset (average dice score \% and average hausdorff distance in mm).  $\dagger$ indicates that the model is pre-trained on ImageNet.}\label{tab1}
\renewcommand\arraystretch{1.2}
\tabcolsep3pt
\scriptsize

\begin{tabular}{c|cc|cccccccc}

\hline
Methods    & DSC$\uparrow$   & HD$\downarrow$    & Aot & Gal    & Kid (L) & Kid (R) & Liv & Pan & Spl & Sto \\ \hline
Swin-Unet$^\dagger$~\cite{swin-unet}  & 79.13          & 21.55         & 85.47          & 66.53          & 83.28          & 79.61          & 94.29          & 56.58          & 90.66          & 76.60          \\
TransUNet$^\dagger$\cite{transunet}  & 77.48          & 31.69         & 87.23          & 63.13          & 81.87          & 77.02          & 94.08          & 55.86          & 85.08          & 75.62          \\
LeViT-Unet$^\dagger$~\cite{levit}  & 78.53          & 16.84         & 87.33          & 62.23          & 84.61          & 80.25          & 93.11          & 59.07          & 88.86          & 72.76          \\
MISSFormer~\cite{missformer} & 81.96          & 18.20         & 86.99          & 68.65          & 85.21          & 82.00          & 94.41          & 65.67          & 91.92          & 80.81          \\  
CoTr~\cite{cotr}       & 86.33          & 12.63         & 92.10          & \textbf{81.47} & 85.33          & 86.41          & 96.87          & 80.20          & 92.21          & 76.08          \\
nnFormer$^\dagger$~\cite{nnformer}    & 86.45          & 14.63         & 89.06          & 78.19          & \textbf{87.53} & 87.09          & 95.43          & 81.92          & 89.84          & 82.58          \\
nnU-Net~\cite{nnunet}    & 87.75          & 9.83          & \textbf{92.83} & 80.66          & 84.86          & 89.78          & \textbf{97.17} & 82.00          & \textbf{92.39} & 82.31          \\ 
UNETR\cite{unetr}     & 79.42          & 29.27         & 88.92          & 69.80          & 81.38          & 79.71          & 94.28          & 58.93          & 86.14          & 76.22          \\
Swin UNETR\cite{swin-unetr} & 85.78          & 17.75         & 92.78          & 76.55          & 85.25          & 89.12          & 96.91          & 77.22          & 88.70          & 79.72          
\\\hline
PHTrans    & \textbf{88.55} & \textbf{8.68} & 92.54          & 80.89          & 85.25          & \textbf{91.30} & 97.04          & \textbf{83.42} & 91.20          & \textbf{86.75} \\ \hline
\end{tabular}
\end{table}

\begin{table*}[b]
\begin{floatrow}
\capbtabbox{
\centering
\renewcommand\arraystretch{1.2}
\tabcolsep2.3pt
\scriptsize

\begin{tabular}{c|c|ccc}
\hline
Methods    & DSC            & RV             & MLV            & LVC            \\ \hline
Swin-Unet$^\dagger$~\cite{swin-unet}  & 90.00          & 88.55          & 85.62          & \textbf{95.83} \\
TransUNet$^\dagger$\cite{transunet}  & 89.71          & 88.86          & 84.53          & 95.73          \\
LeViT-Unet$^\dagger$~\cite{levit} & 90.32          & 89.55          & 87.64          & 93.76          \\
MISSFormer~\cite{missformer} & 90.86          & 89.55          & 88.04          & 94.99          \\
nnFormer$^\dagger$~\cite{nnformer}   & 91.62          & \textbf{90.27} & 89.23          & 95.36          \\
nnU-Net~\cite{nnunet}    & 91.36          & 90.11          & 88.75          & 95.23          \\ \hline
PHTrans    & \textbf{91.79} & 90.13          & \textbf{89.48} & 95.76          \\ \hline
\end{tabular}
}{
\caption{Results of segmentation on the ACDC dataset. $\dagger$ indicates that the model is pre-trained on ImageNet.}
\label{tab2}
}
\capbtabbox{
\centering
\scriptsize
\renewcommand\arraystretch{1.2}
\tabcolsep4pt
\begin{tabular}{c|cc}
\hline
Methods  & Params (M) & FLOPs(G) \\\hline
CoTr~\cite{cotr}     & 41.9       & 318.5    \\
nnformer~\cite{nnformer} & 158.7      & 171.9   \\
nnU-Net~\cite{nnunet}  & \textbf{30.8}       & 313.2    \\
UNETR~\cite{unetr}  &  92.6       & \textbf{82.6}     \\
Swin UNETR~\cite{swin-unetr}  &  61.98      & 394.84    \\
PHTrans  & 36.3        & 187.4  \\ \hline  
\end{tabular}
}{
 \caption{Comparison of number of parameters and FLOPs for 3D segmentation models in BCV experiments.}
 \label{tab3}
}
\end{floatrow}
\end{table*}

\subsubsection{Comparing with State-of-the-arts.}

\begin{figure}[t]
\centering
 \includegraphics[scale=0.48]{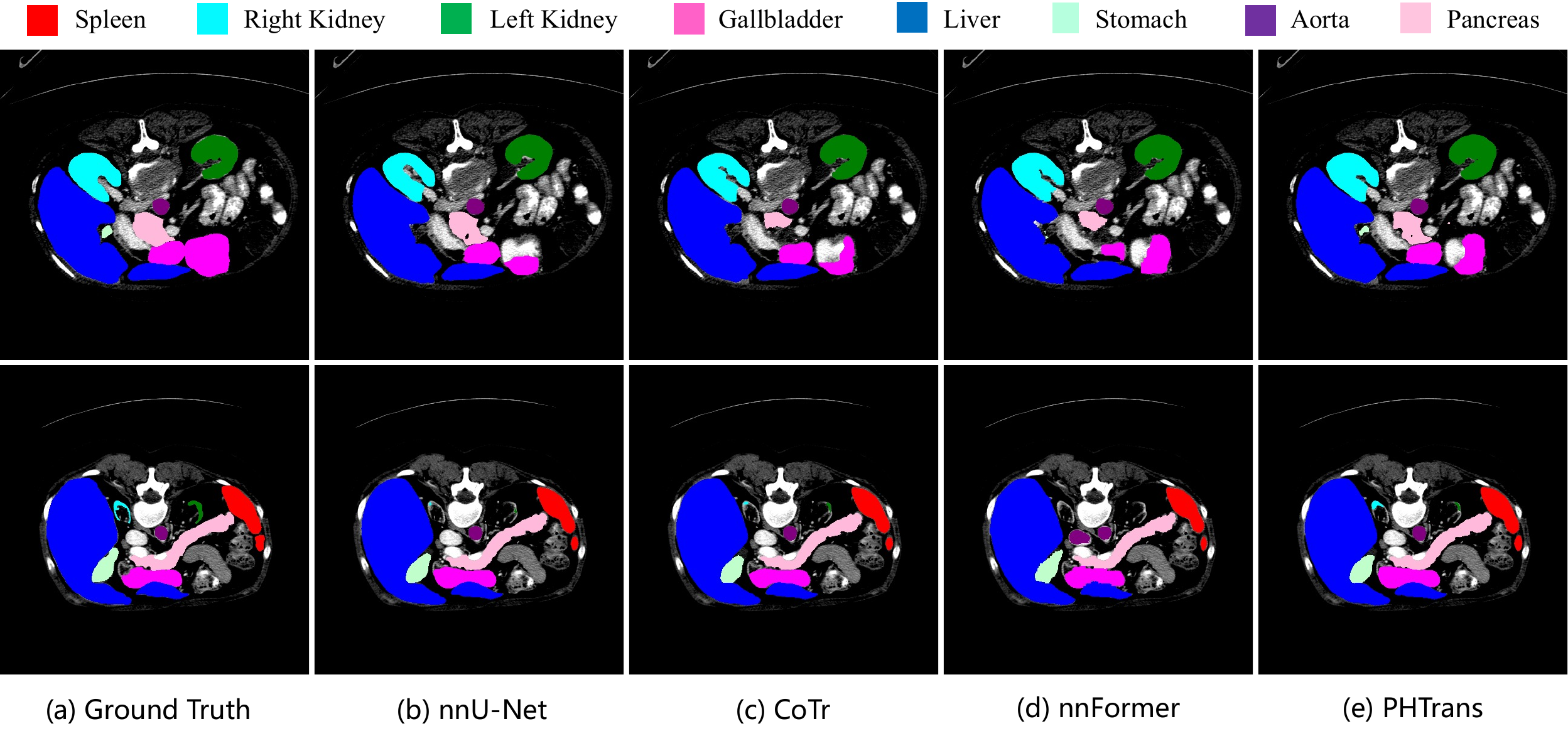}
\caption{Qualitative visualizations of the proposed PHTrans and other 3D methods. } \label{fig3}
\end{figure}

We compared the performance of PHTrans with previous state-of-the-art methods. In addition to the hybrid architectures mentioned in the introduction, it also includes, LeVIT-Unet~\cite{levit}, Swin-Unet~\cite{swin-unet}, and nnU-Net~\cite{nnunet}. Furthermore, we reproduced Swin UNETR\cite{swin-unetr} by modifying ViT in the encoding stage of UNETR\cite{unetr} into a Swin transformer and also evaluated the performance of UNETR and Swin UNETR in the same way and used the same dataset partition as ours. The results of segmentation on the BCV dataset are shown in Table \ref{tab1}. Our PHTrans achieves the best performance with 88.55\% (DSC$\uparrow$) and 8.68 (HD$\downarrow$) and surpasses the previous best model by 0.8\% on average DSC and 1.15 on HD. The representative samples in Fig.~\ref{fig3} demonstrate the success of identifying organ details by PHTrans, e.g., ``Stomach" in rows 1, 2, and ``Left Kidney" in rows 2. The results of segmentation on the ACDC dataset are presented in Table~\ref{tab2}. Similarly, it is obviously apparent that PHTrans is competitive with other state-of-the-art methods by achieving the highest average DSC. It is worth mentioning that Swin-Unet, TransUNet, LeViT, and nnFormer use the pre-trained weights on ImageNet to initialize their networks, while PHTrans was trained on both datasets from scratch. Additionally, we compared the number of parameters and FLOPs to evaluate the model complexity of 3D approaches, i.e., nnformer, CoTr, nnU-Net, UNETR, Swin UNETR and PHTrans, in BCV experiments. As shown in Table~\ref{tab3}, PHTrans has few parameters (36.3M), and its FLOPs (187.4G) is significantly lower than CoTr, nnU-Net and Swin UNETR. In summary, the results of PHTrans on BCV and ACDC datasets fully demonstrate its excellent medical image segmentation and generalization ability with preserved moderate model complexity.

\begin{table}[tbp]
\centering
\caption{Ablation study on the architecture. 
}\label{tab4}
\renewcommand\arraystretch{1.2}
\tabcolsep2.5pt
\scriptsize
\begin{tabular}{c|cc|cccccccc}

\hline
Methods    & DSC$\uparrow$   & HD$\downarrow$    & Aot & Gal    & Kid (L) & Kid (R) & Liv & Pan & Spl & Sto \\
\hline
3D Swin-Unet      & 83.47 & 20.13 & 87.46 & 71.37       & 86.77      & 83.53      & 95.47 & 72.51    & 92.19  & 78.48   \\
3D Swin-Unet+PCM & 84.95 & 19.52 & 87.36 & 77.55       & 85.76      & 84.48      & 95.59 & 79.66    & 88.93  & 80.26   \\
PHTrans w/o ST   & 87.71 & 14.37 & 92.22 & 79.35       & 85.94      & 88.08      & 96.62 & 83.26    & 90.63  & 85.59   \\
PHTrans w/o PCM  & 86.10          & 18.29         & 88.73          & 76.56          & \textbf{86.81} & 85.56          & 96.09          & 78.04          & \textbf{94.28} & 82.77          \\
PHTrans          & \textbf{88.55} & \textbf{8.68} & \textbf{92.54} & \textbf{80.89} & 85.25          & \textbf{91.30} & \textbf{97.04} & \textbf{83.42} & 91.20          & \textbf{86.75}  \\ \hline
\end{tabular}
\end{table}

\subsubsection{Ablation Study.}

Using the modified 3D Swin-Unet as the baseline, we progressively integrated the components of PHTrans to explore the influence of different components on the model performance. Table~\ref{tab4} provides quantitative results of the ablation study on the architecture. ``+PCM" denotes using stacked pure convolutional modules instead of a strided convolution operation for patch partition, while ``w/o PCM denotes the opposite. ``w/o ST" means that the parallel hybrid module in PHTrans removes the Swin Transformer blocks, resulting in a similar architecture to nnU-Net. From these results, it can be seen that the performance of 3D Swin-Unet and PHTrans is improved by PCM, which is owed to its ability to capture fine-grained details in the first few stages. Furthermore, the PHTrans brings more significant performance gains compared with the single architecture and outperforms ``3D Swin-Unet$+$PCM" and ``PHTrans w/o ST" by 3.6\% and 0.84\% in average DSC and 10.84 and 5.69 in HD, respectively. The results indicate the effectiveness of using a parallel combining strategy of CNN and Transformer to aggregate global and local representations.

\subsubsection{Discussion.}

In PHTrans, a vanilla Swin Transformer and simple convolutional blocks are applied, which demonstrates that significant performance gains stem from the parallel hybrid architecture design rather than the Transformer and CNN blocks compared to state-of-the-arts. Furthermore, PHTrans is not pre-trained since there is no large enough all-purpose dataset of 3D medical images so far. From the above considerations, in the future, we will elaborately design Transformer and CNN blocks and explore how to pre-train Transformer end-to-end to further improve the segmentation performance.

\section{Conclusions}
In this paper, we propose a parallel hybrid architecture (PHTrans) based on the Swin Transformer and CNNs for accurate medical image segmentation. Different from other hybrid architectures that embed modular Transformer into CNNs, PHTrans constructs hybrid modules consisting of Swin Transformer and CNN throughout the model, which continuously aggregates hierarchical representations from global and local features to give full play to the superiority of both. Extensive experiments on BCV and ACDC datasets show our method is superior to several state-of-the-art alternatives. As an all-purpose architecture, PHTrans is flexible and can be replaced with off-the-shelf convolution and Transformer blocks, which open up new possibilities for more downstream medical image tasks. 

\subsubsection{Acknowledgment.}This work was supported in part by the National Key R\&D Program of China (No.2018AAA0102600) and National Natural Science Foundation of China (No.62002082).

%

%
%
%
\bibliographystyle{splncs04}
\bibliography{refer}
%

\end{document}